# Towards an Ontology of Traceable Impact Management in the Food Supply Chain

Bart Gajderowicz[1], Mark S Fox[2], Yongchao Gao[2]

[1] bartg@mie.utoronto.ca
Urban Data Centre
School of Cities
University of Toronto
Toronto, Canada

[2] msf@mie.utoronto.ca
Urban Data Centre
School of Cities
University of Toronto
Toronto, Canada

[3] gaoyc@sieti.cn
Key Laboratory of Computing Power
Network and Information Security
Shandong Computer Science Center
Qilu University of Technology
(Shandong Academy of Sciences)
Jinan, China

**Abstract**
The pursuit of quality improvements and accountability in the food supply chains, especially how they relate to food-related outcomes, such as hunger, has become increasingly vital, necessitating a comprehensive approach that encompasses product quality and its impact on various stakeholders and their communities. Such an approach offers numerous benefits in increasing product quality and eliminating superfluous measurements while appraising and alleviating the broader societal and environmental repercussions. A traceable impact management model (TIMM) provides an impact structure and a reporting mechanism that identifies each stakeholder's role in the total impact of food production and consumption stages.

The model aims to increase traceability's utility in understanding the impact of changes on communities affected by food production and consumption, aligning with current and future government requirements, and addressing the needs of communities and consumers. This holistic approach is further supported by an ontological model that forms the logical foundation and a unified terminology. By proposing a holistic and integrated solution across multiple stakeholders, the model emphasizes quality and the extensive impact of championing accountability, sustainability, and responsible practices with global traceability.

With these combined efforts, the food supply chain moves toward a global tracking and tracing process that not only ensures product quality but also addresses its impact on a broader scale, fostering accountability, sustainability, and responsible food production and consumption.

## Introduction

United Nations Sustainability Goal (UNSDG) number 2, "Zero Hunger", shifts the focus from quality and performance metrics of food production to understanding how food supply chains impact stakeholders who suffer from hunger and food insecurity. "Food security requires a multi-dimensional approach – from social protection to safeguard safe and nutritious food, especially for children, to transforming food systems to achieve a more inclusive and sustainable world. There will need to be investments in rural and urban areas and in social protection so poor people have access to food and can improve their livelihoods."[1] This shift in focus necessitates a-comprehensive understanding of key stakeholders, such as the poor, those living in remote areas or where access to food is limited, as well as organizations that fund and operationalize food production and distribution on these consumer stakeholders. It also focuses on the impact the supply chain has on their desired outcomes, whether not being hungry, eating notional food, and delivering food that is of high quality, safe to consume, in a secure manner that abides by different jurisdictions the food may go through. It requires the ability to trace how activities and events in the supply chain affect stakeholder outcomes.

The pursuit of traceable impact within the agricultural food chain is a multifaceted endeavour, as outlined by Pizzuti and Mirabelli (2003). It seeks to establish a global standard for impact tracking and tracing processes. This initiative begins with a thorough analysis of the food supply chain, dissecting the intricate web of production, processing, distribution, and consumption. Building upon this analysis, it extends into the realm of supply chain modelling, laying out the structural and operational dynamics that govern the flow of food products. Integral to this pursuit is the development of an impact management model, which aims to determine the effects of the food supply chain on both socio-economic and environmental fronts. Underpinning these models is the critical role of data collection and data modelling, which provide the empirical basis for understanding and improving the supply chain. Finally, the process culminates in creating software tools designed to enhance food chain processes, fostering positive impacts and mitigating risks. This comprehensive approach addresses the complexities of the global food system and sets a trajectory for sustainable and responsible food production and distribution.

This paper proposes an ontology-based modelling and analysis framework that integrates impact management and traceability within the food supply chain. In the next section, we discuss traceability in the Supply Chain. We then introduce the concept of Traceable Impact Management. We review the areas of Impact Management, Traceability in the supply chain and the benefits of combining the two. We next define the Traceable Impact Management ontology, which combines and extends the ISO/IEC 5087-1 smart city standard and the Common Impact Data Standard (CIDS), that form the foundation of our framework.

## Traceability in the Supply Chain

Traceability, as it pertains to the food supply chain, is the ability to track each step of the food production, processing, transportation and consumption processes and the ability to retrace each tracked step from one point to its origins (Olsen & Borit, 2013); Mc Carthy et al., 2018). It entails knowing the atomic events that transform food products from raw materials into packaged goods, the consumption of food products, as well as the by-products and waste each activity produces. Traceability also entails the who dimension, tracking the role of each stakeholder and the activities they perform in transforming the products through production, processing and consumption. The tracked steps should have a temporal dimension, capturing when and for how long that activities or events occurred. Traceability should have a geospatial component, capturing where the tracked events occurred. The tracked steps also should include the quantity dimension, knowing what and how much of it was produced, etc. Finally, traceability should capture the quality of products to trace and identify the origins of issues in the supply chain.

---

[1] https://www.un.org/sustainabledevelopment/hunger/

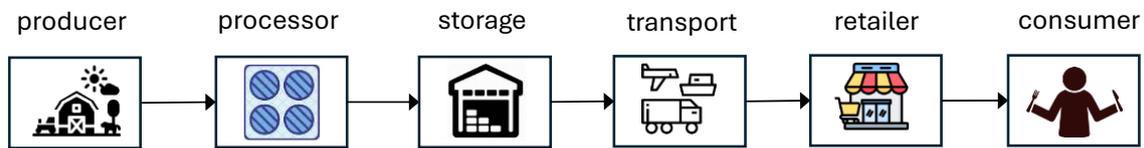

*Figure 1. Beef food supply chain use case.*

There are four pillars of food security, namely availability, access, utilization, and stability (McCarthy et al., 2018). Consider the supply chain through which beef moves from production to consumption, as illustrated in Figure 1. The linear nature of this process hides the complexities required to achieve accountability, sustainability, and traceability and the role they play in food insecurity.

During beef production, raising cattle is susceptible to many factors, including the environment, viruses, and the availability of quality feed and medicines (Roberts et al., 2004). During environmental issues, such as drought or disease, beef production will slow down at best. At worst, it may contaminate other parts of the supply chain or even reach the market. Next, beef is processed by combining ingredients and other materials to produce products like packaged beef patties ready for distribution. The transfer and storage of raw beef are susceptible to many airborne contaminants and temperature fluctuations, increasing the opportunity for contamination and spoilage. The equipment transforming cattle into beef and packaged patties requires resources, raw materials, and personnel to operate and maintain. The personnel working at the facilities are reliant on the region where they live, such as housing, health care, education, and other social services. Any outbreaks of disease among the personnel or any issues in their lives that impact their productivity will slow down production or affect quality.

The distribution network has a broader point of contact and is multi-spatial. For example, storing and transporting beef requires many points of transferring the product from one organization to another, often crossing regions with different food handling policies and jurisdictions. Political and regulatory factors may contribute to food insecurity. Food may not be delivered where regulations have higher standards. Policy or social demands may cause food to be processed differently than expected. Wrong or incompatible definitions of food between stakeholders or across borders make it difficult to track food transformation consistently. Economic factors, such as currency fluctuation, may impact food delivery to regions where lower profits are due to lower prices or remote areas where delivery costs are higher (Carney, 2011).

Finally, many compounding factors of food insecurity can only be detected at the final point of delivery when selling or consuming food (Hoek et al., 2021). Prematurely expired meat due to long deliveries or broken-down refrigeration units makes the food inedible. Wrong or contaminated ingredients may cause food to lose its nutritional value.

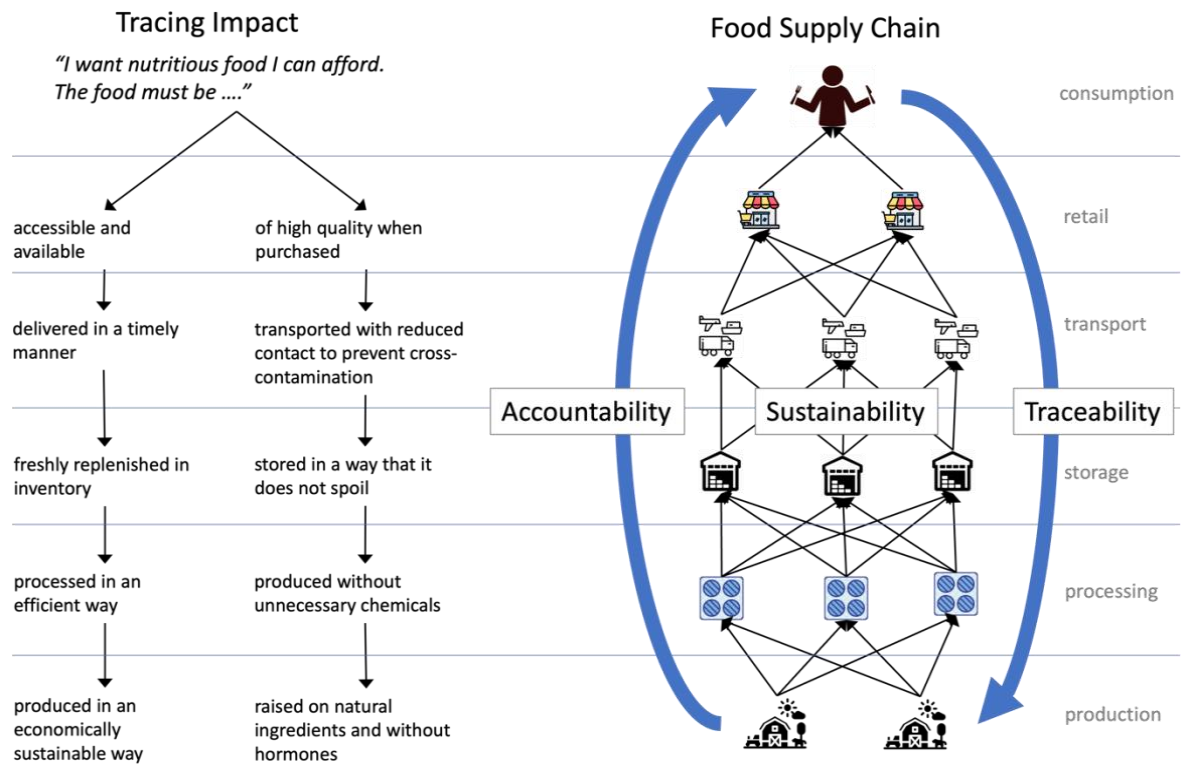

*Figure 2. Beef food supply chain with impact, accountability, sustainability, and traceability.*

Figure 2 illustrates how food insecurity factors impacting the consumer (nutrition and affordability) are propagated to the producer. To achieve accountability, we must begin at the start of the food supply chain, where the raw materials are produced, and continue through to the final stage of consumption. To achieve sustainability, we must have a complex and interconnected network of stakeholders. Hence, rather than relying on linear processes, sustainability is achieved through the availability of different paths goods can travel through. Finally, to achieve holistic traceability, we must ensure that every point in the path is accounted for.

Traceability within the supply chain forms the basis for efficient recall procedures, minimizing losses in case of quality concerns. Moreover, it provides valuable information about raw materials, facilitating better quality and process control while eliminating redundant measurements in successive steps. It fosters an incentive to maintain the inherent quality of raw materials. It aligns with governmental requirements, such as confirming the country of origin and each step of the food production process, production, packaging, transfer, storage, and sale. Furthermore, traceability systems help eliminate redundant measurements and inspections by ensuring that information gathered at one stage of the supply chain is accurately relayed and utilized downstream, enhancing overall efficiency.

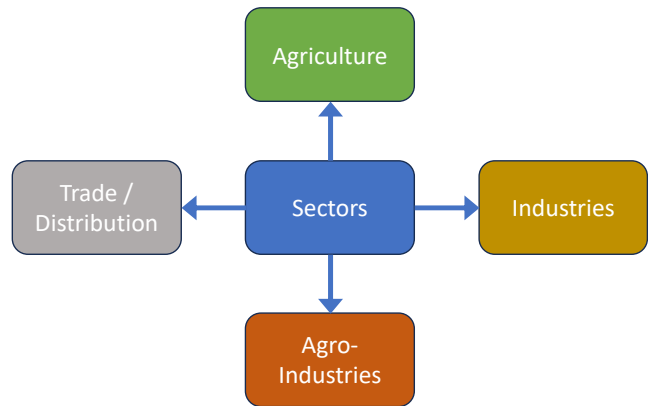

*Figure 3. Food supply chain sectors.*

Four key economic sectors make up the food supply chain and its stakeholders, as illustrated in Figure 3, namely:
- Agriculture produces and processes raw materials, distinguished into segments like crop cultivation processes, aquaculture, livestock or poultry production processes.
- Agro-industries include a variety of processes for processing, combining, packaging and testing food products.
- Trade/Distribution sectors perform activities such as buying, transporting, warehousing and selling food products.
- Industries are auxiliary to the other three sectors by providing technical equipment and supplies.

To successfully create a traceable impact model, we must identify all stakeholders within these sectors. Such stakeholders may include food producers, processors, logistics companies, commercial organizations, and customers. They may include consolidators such as those that buy bundles or in bulk for resale.

Food Supply Chain Modeling employs a variety of methodologies to develop a trace graph, which is essential for visualizing and optimizing the flow of goods from producers to consumers. Among the methodologies utilized are Petri-nets (Balamurugan et al., 2021), which offer a mathematical modelling language ideal for describing distributed systems; and Integration Definition (IDEF) (Thakur et al., 2009), a family of modelling languages in the field of systems and software engineering. Event-Driven Process Chains (EPC) (Bevilacqua et al., 2009) also play a crucial role in depicting business process workflows, providing clarity on event sequences and organizational functions. Furthermore, modelling approaches such as UML activity diagrams (Islam at el., 2021) and Business Process Modeling (BPM) (Gayialis, 2022; Pizzutti, 2012) has significantly advanced these efforts, with various initiatives and notations emerging to standardize and improve the representation and analysis of business processes within the food supply chain, thereby enhancing traceability and efficiency. Additional methods have been identified as having the potential to model a traceable food supply chain (Pizzuti, 2015; Verdouw, 2010), including Structured Analysis and Design Technique (SADT), which focuses on the systematic analysis of information flow; Electronic Document Management (EDOC) for Business Processes, and Activity Decision Flow (ADF) contribute detailed perspectives on operational activities and decision-making processes.

In modelling the food supply chain, it is important to distinguish between what is planned versus what has happened. The former requires specifying the network of activities, resources, stakeholders, goals, etc., while the latter extends it by capturing events that occur across the supply chain, necessitating the collection of data throughout. Data collection encompasses a wide array of traceable elements crucial for maintaining transparency and ensuring accountability throughout the chain. Traceable data include detailed records of the primary stakeholders producing and distributing food products. This data encapsulates the responsibilities and liabilities associated with the product and its origin. Additionally, data regarding the product are collected, including information on the product's state, the machinery and utensils used in its production, and any food additives, feed, or fertilizers employed in the process. Just as important, it requires data about the product as it moves through the supply chain. For example, the transportation of meat requires capturing product temperatures over time. Central to tracing is product identification, which is achieved through unique identifiers, such as Radio Frequency Identification (RFID) and Near Field Communication (NFC) technologies, and the reuse of existing standards such as GS1 and EPC.

Traceability within the food supply chain can be conceptualized at various levels of abstraction. Internal or local traceability refers to the detailed monitoring within an organization's own system. It includes meticulous tracking of product identification, resources such as feed, seeds, and herbicides, along with packaging material and equipment. Furthermore, it records storage management and processing times, providing a comprehensive view of the product's journey within a single entity's domain. This level of traceability is instrumental in ensuring the integrity of products and processes within a particular actor's scope of the supply chain.

Conversely, global traceability provides a macroscopic view of the product's journey across different stakeholders in the system. This level of traceability focuses on less granular but equally essential data points such as arrival time, transport duration, the origin of the product, and, particularly in the case of meat products, the identification of the animal or group from which the meat was sourced. Such information is vital for broader oversight and coordination among various supply chain actors, enabling tracking products across different stages and entities, from farm to table.

## Traceable Impact Management

Impact "refers to the *intended* and *unintended* (*positive or otherwise*) changes (*outputs, outcomes*) that occur across the organization (within and/or across its programs) and with its *stakeholders* (including users, clients, partners, etc.) over a period of time (short term, long term) *as a result of the organization's activities*"[2] Numerous impact models, such as the Logic Model, Theory of Change, Outcome Map, Outcome Chain, and Impact Map, offer various lenses through which impact can be articulated and measured. Despite their differences, these models share core concepts, allowing the formulation of a unified model that encapsulates diverse perspectives without favouring any single one (Ruff, 2021). The

---

[2] https://innoweave.ca/en/modules/impact-measurement

Impact Management Project (IMP) [3] has refined, through a consensus of over 2000 practitioners, a multi-dimensional approach to modelling and measuring impact. This includes the dimensions of What, Who, How Much, Contribution, and Risk, which together provide a comprehensive framework for understanding the changes brought about by social purpose organizations.

*Table 1. Six dimensions of the Common Impact Data Standard.*

| Impact Dimension | Impact question each dimension seeks to answer |
|---|---|
| What | What outcome is occurring in the period? |
| Who | Who experiences the outcome? Who should be? |
| How much | How much of the outcome is occurring – across scale, depth, and duration? |
| Contribution | How big of a contribution the outcome makes? |
| Risk | What is the risk to people and the planet that impact does not occur as expected? |
| How | The processes by which an organization delivers outcomes to its stakeholders. |

The five dimensions were extended with the development of the Common Impact Data Standard (CIDS) (Table 1) where the introduction of the "How" dimension defines the steps for achieving outcomes (Fox and Ruff, 2021). CIDS emphasizes the need to detail the nature and significance of outcomes, the characteristics and needs of those affected, the extent of change attributed to the organization's interventions, and the risks involved should the anticipated changes not materialize. This framework applies to all types of organizations, not only those with a social purpose, by providing a structured approach to documenting and evaluating the impact of their activities.

Impact Management in the food supply chain entails establishing and subsequent monitoring of an Impact Model specific to the food supply chain. This model requires comprehensive definitions for key concepts (Fox and Ruff, 2021), including:
- outcomes (e.g., intended and unintended changes due to the food supply chain's production and consumption phases),
- stakeholders (e.g., those expressing a desire for or inadvertently feeling the impact of outcomes), and
- indicators (used to measure and report on the impact of the supply chain on stakeholder outcomes.

Traceable impact management links stakeholder outcomes to traceable resources, activities and events in the supply chain. It requires a detailed analysis of how activities and events, ranging from raw material sourcing, production, and transportation to consumer practices, are linked, causally or otherwise, to positive and negative stakeholder outcomes.
Yadav (2022) asserts that the scope of impact tracing extends well beyond the mere production, use, or consumption of goods and services. It involves a nuanced exploration of the repercussions of these activities. Modelling the food supply chain transcends the traditional confines of logistical networks and economic transactions. It must also factor in the resultant outcomes, such as food waste, food safety, security, and sustainability integration within the chain. These considerations are pivotal in crafting a supply chain that is efficient, economically viable, socially responsible, and environmentally sustainable, addressing the pressing challenges of our time.

---

[3] https://www.theimpactprogramme.org.uk/portfolio/impact-management-project/

Understanding the social implications of food production and consumption is integral to developing sustainable food systems. Combining food traceability with impact management enables stakeholders to discern which modifications in the supply chain can result in the most significant benefits for communities affected by these activities. For example, identifying how changes in farming practices can lead to better environmental outcomes while identifying improvements in distribution can enhance food security in vulnerable populations. These insights drive responsible decision-making and foster community-centric approaches. Moreover, the inherent quality of raw materials is given more emphasis, with traceability acting as a catalyst for maintaining high standards. Producers have a heightened incentive to ensure the quality of their goods from the outset, knowing that the end-to-end visibility of their products can influence consumer trust and brand reputation.

Furthermore, establishing efficient extensions of the utility of traceability to the impact of changes on communities affected by food production and consumption aligns with current and future government requirements and addresses the needs of communities and consumers. For example, when a product is identified as harmful or non-compliant with quality standards, a traceability system can rapidly locate the product's journey through the supply chain, enabling targeted recalls. When the impact of quality issues is further down the supply chain or impacts stakeholders removed from the supply chain (e.g. distributed or remote resellers, communities along the transportation path), it is more difficult to minimize economic losses and protect consumer health across such dispersed stakeholders without identifying impacted stakeholders and their outcomes. Additionally, this granular level of information about raw materials and production processes can significantly enhance quality and process control. Producers can optimize their processes by understanding the provenance and handling of raw materials and the quality issues have on local communities, including their local workforce as well as a local and remote consumer base.

Finally, the strategic implementation of traceability and impact management systems in the food supply chain aligns with the evolving regulatory landscape. These systems are instrumental in complying with current and forthcoming government mandates, such as confirming the country of origin or adhering to sustainability guidelines. Not only does this ensure legal compliance, but it also opens new marketing opportunities. Producers can now differentiate their products based on special features of raw materials or unique production attributes, catering to niche markets and consumer trends that value transparency and ethical production. As regulations become more stringent and consumers more discerning, traceability and impact management become tools for risk mitigation, market differentiation and competitive advantage.

## Ontologies for Tracing Impact in the Supply Chain

In this section we review three ontologies that form the basis for the Traceable Impact Management ontology defined in the next section:

1. The ISO/IEC 5087 series of data standards and the TOVE Enterprise Ontology provide representations of time, resources, activities, states, and organizational structure and behaviour.
2. The quality/traceability ontology portion of the TOVE ontology, which provides representations for how resources are traced through the activities they participate in.
3. The Common Impact Data Standard ontology for impact modelling, which provides representations for stakeholders, outcomes, indicators and reporting.

## Activity Ontology (5087-1)

The ISO/IEC 5087 is a series of standards for smart city data. ISO/IEC 5087-1:2023 is the first in the series and provides a standard for foundation level concepts such as: activities, resources, agents, time, geospatial, and organization structure.

## 5087-1 Activity Pattern

To sufficiently trace supply chain activities, 5087-1 provides the Activities-State model, based on the TOVE activity model (Fox, 1992; Fox et al., 1993) (Figure 4), which includes the following components:
- **Activity**: An action is represented by the combination of an activity and its corresponding enabling and cased states.
- **State**: Describes what is true at some time t, for example, what is true prior to the performance of an activity and what is true after its performance.
- **Causality**: What conditions (represented as states) have to be satisfied (true) to enable the performance of an activity? What conditions (state) will be satisfied (i.e., caused by) when the activity has been performed?

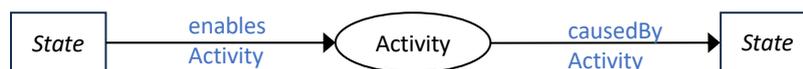

*Figure 4. Activity-State Model.*

## TOVE: Activity Abstractions

Next, we may store information about activities and different levels of abstraction. The Activity Abstraction (Figure 5), based on the TOVE ontology (Fox, 1992; Fox et al., 1993), deconstructs complex activities into levels of abstraction. This includes the sub-activities the activity can be divided into, and the super-activities the activity is a part of.

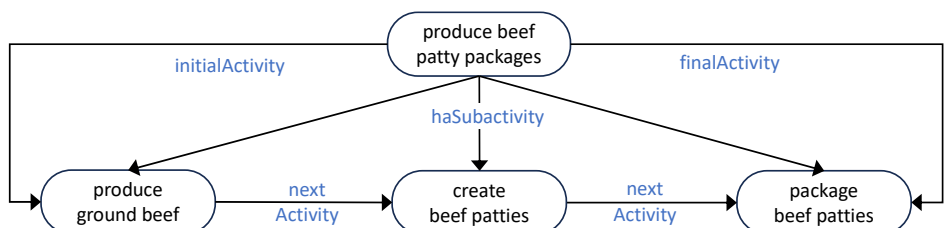

*Figure 5. Activity Abstraction.*

Figure 6 contains an activity, along with its enabling and caused states, called an **activity-state cluster**. An activity-state cluster includes the enabling states (*es:produce ground beef*) that specify what has to be true in order for the activity cluster to be performed. It also includes the caused states (*post:produce ground beef*) of the activity cluster. This defines what will be true of the world once the activity has been completed. The states can be organized as a tree of states that form a conjunction and/or disjunction of enabling states.

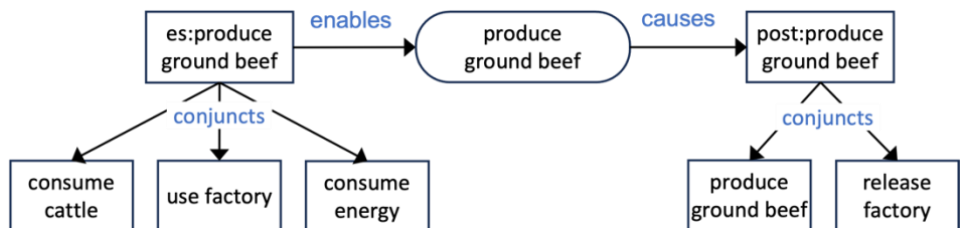

*Figure 6. Activity State Cluster.*

## 5087-1 Time Ontology

The time ontology in 5087-1 is based on OWL-Time (Cox & Little, 2022; Hobbs & Pan, 2006; Allen, 1984). It defines the temporal aspects of activities through the constructs of points, periods, and relations. A time-point, signifying an atomic instance, exists within a broader interval called a time-interval. A time-interval is demarcated by boundary time-points that encapsulate the duration of an activity, defined by its start and end time-points. The concept of a time window further refines this by establishing the earliest and latest thresholds for the start and end of an activity, introducing flexibility and constraints within temporal planning. Finally, time interval relations (Figure 7) articulate the sequential or overlapping nature of intervals. When time-intervals are associated with activities one can determine whether one activity precedes, follows, or coincides with another. That same applies to states. This structured representation of time is pivotal in organizing, analyzing, and optimizing activities within various applications.

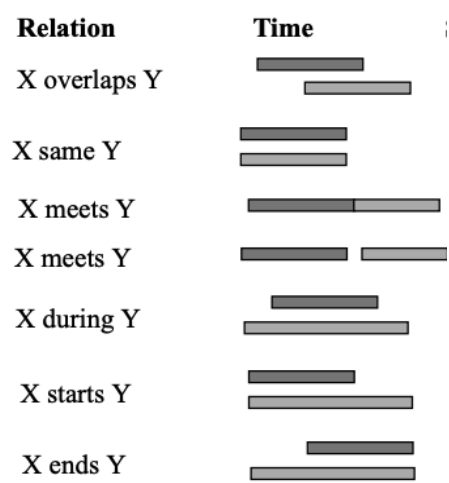

*Figure 7. Time Interval Relations.*

## 5087-1 Resource Ontology

The 5087-1 **resource** taxonomy, based on TOVE (Fox et al., 1993; Fadel et al., 1994), in Figure 8, captures basic properties such as quantity, unit of measure, location and time. Together, they convey how much of a resource is available at some point in time at some location. Secondly, it captures the allocation, past, present and future, to activities that either use or consume it. Finally, it captures contextual properties, such as whether a resource can be subdivided and still useable by an activity. The properties are important for tasks such as resource scheduling, allocation, and traceability. For example, knowing what processes and resources were involved in creating a beef patty package (ground beef and spices) allows for the traceability of resources used to make that package. In this way, we can track each process and the impact the consumption of resources has on the creation of each resource.

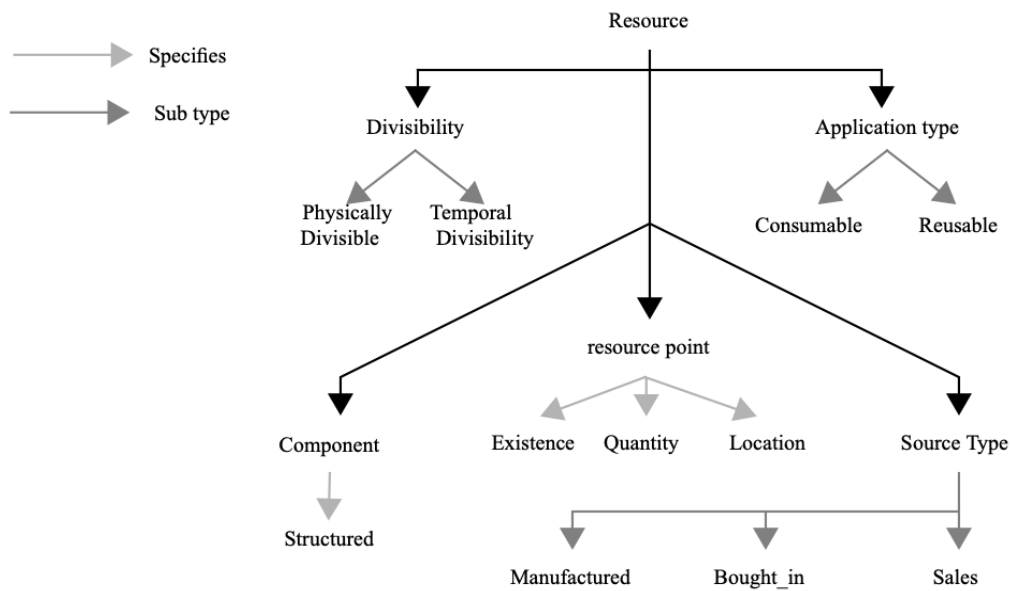

*Figure 8. TOVE Resource Taxonomy.*

Application types are analysed to understand if and how an activity uses and releases a resource or whether the resource is depleted, which in turn influences replenishment cycles and cost implications. Divisibility plays a key role in determining a resource's utility, dictating whether it can be allocated partly to concurrent activities without losing functionality. For example, a batch of ground beef can be split into multiple beef patty packages, but a single patty box cannot be divided, instead being consumed (i.e., bought/sold) in its entirety. Resources are not limited to the raw materials used to create food products. For example, the machinery, energy, and human capital used to transform raw beef into beef patties must be tracked, and their impact must be evaluated. For example, machines use energy, which impacts the environment. Human capital is used, which has a positive impact on the local community through employment, tax income, and so on.

To act on traceable impact metrics, utilizing resources to be shared among various activities is fundamental to optimizing its use across a system. For example, the utilization of human capital to process meats impacts local communities from which the workers are from. Transferring beef patties impacts the local environment through which it is transported, taking up roads and railways while refuelling at gas stations. The selling and consumption of beef patties have an economic and health-related impact on the communities where consumers live. Spatial considerations are addressed by the resource's location, which impacts logistics, accessibility, and response times. Lastly, the commitment of a resource to activities at a specific time encapsulates its planned utilization, reflecting its role and importance within the operational framework. These facets collectively inform the strategic management of resources, ensuring their effective and efficient use in complex systems.

## TOVE Traceability Ontology

The TOVE Quality Ontology (Kim et al., 1995) provides a formal representation of quality that transcends specific domains by employing first-order logic to define terms, relationships, and axioms. Predicated on the notion that quality equates to "conformance to requirements," the ontology divides the domain of quality into the distinct sub-domains of measurement, analysis, identification, and traceability. The Identification Ontology allows for unique identification and classification of entities, ensuring that the quality of one can be distinguished and compared to another. The Traceability Ontology enables the determination of all pertinent entities and their attributes that influence the quality of a specified entity within an enterprise can be traced and identified. This approach establishes a foundational capability for traceability to analyze and address quality-related issues.

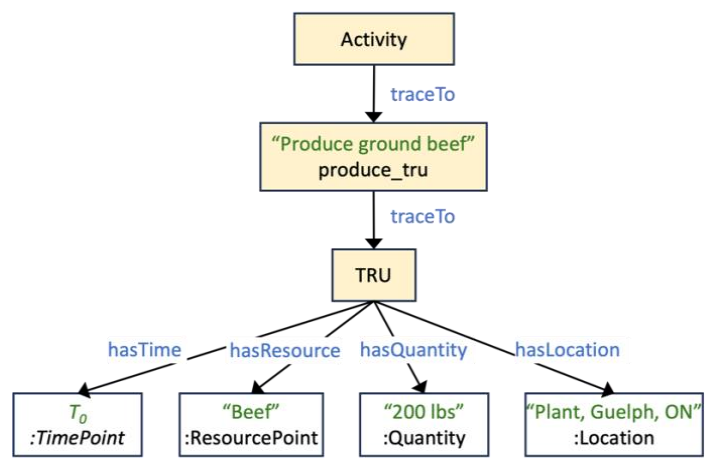

*Figure 9. Traceable Resource Unit (TRU) model example of a "produce_tru" for "Produce ground beef".*

The traceability ontology defines a Traceable Resource Unit (TRU) (Figure 9) as a homogeneous collection of resources categorized by a class and quantified by its interaction with a primitive activity (Kim et al, 1995). A TRU identifies the state of a resource at a given point in time and location, hence a TRU's existence is recognized at the point of the resource's first use, consumption, production, or release, with its quantity determined and fixed at that initial time-point and location. For example, Figure 9 illustrates a TRU for the production of the beef patty resource ("produce ground beef") along with its time point ($T_0$), value, quantity and units of measure (200 lbs), and its location in the meat packing plant ("Plant, Guelph, ON"). Quantitative changes to a TRU are recorded at specific state changes—such as post-production or release—and remain unchanged in the absence of a new transaction. A TRU has no measured quantity before its inception but maintains a measurable quantity, including zero if consumed, after that. This ontology disallows the post-recognition incrementation of a TRU's quantity and asserts that individual units within a TRU are indistinguishable, precluding traceability within a single TRU. Furthermore, suppose a TRU's consumption requires more resources than is available. In that case, it necessitates drawing from additional TRUs, with the consequence that aggregating multiple TRUs does not preserve the identity of any singular TRU.

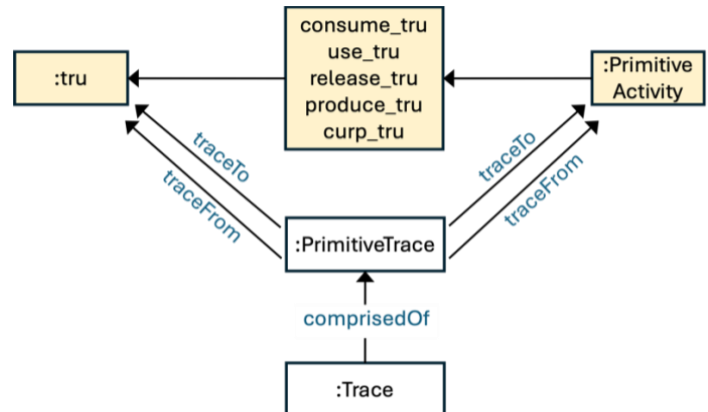

*Figure 10. A Trace instance is comprised of PrimitiveTrace instances that link to and from TRUs and activities.*

Different TRUs and primitive activities can be linked together to form a trace path, as per Figure 10 (Kim et al., 1999). TRUs and activities are connected by either *traceTo* or *traceFrom* a :*PrimitiveTrace* instance. These properties allow one to trace between a TRU and an activity, as well as between TRUs and between activities. The :Trace class allows for the connection of multiple :*PrimitiveTrace* instances, which together comprise a trace path, with connected TRUs and activities.

## Common Impact Data Standard

The Common Impact Data Standard (CIDS), depicted in Figure 11, is a data standard for defining and measuring the impact of activities (Fox and Ruff, 2021). Originally, it was designed to address the historical challenge social purpose organizations (SPOs) face in measuring their social and environmental impacts. It can be generalized to address the difficulty in satisfying the need for uniform measurement methods for benchmarking and the need for flexibility in reporting across domains. Developed as part of the Common Approach to Impact Measurement[4], CIDS offers a standardized framework for representing an SPO's impact model and its effects on stakeholders, combining uniformity and adaptability. This initiative, backed by 48 community partners and funded by the Government of Canada and large private foundations, significantly advances harmonizing impact measurement practices among diverse social organizations.

The benefits of using a standardized and uniform way to represent the impact model and the impact on stakeholders are:

1. **Better impact:** Organizations' greatest impact emerges when data is unified. A shared impact data standard lets networks combine data to enhance and optimize impact.
2. **Sophisticated analysis**. CIDS allows researchers to merge data, enabling diverse studies like longitudinal and transversal. This promotes deeper insights into needs and effective solutions.

---
[4] https://www.commonapproach.org/

3. **More autonomy**. CIDS standardized formats for portfolio-level impacts. SPOs can measure impact tailored to their data needs and those of their partners, donors, investors, and government agencies.
4. **Less paperwork**: A unified impact data model meets the diverse reporting needs of funders, simplifying custom reporting for SPOs.
5. **Greater visibility:** Enabling tagging of an organization's content on the internet makes it easier for search engine users to find impact content on the web.
6. **More versatility:** A common data model simplifies integration of impact measurement with standards like UN SDG Global Indicator Framework, IRIS+, and IATI Standard.

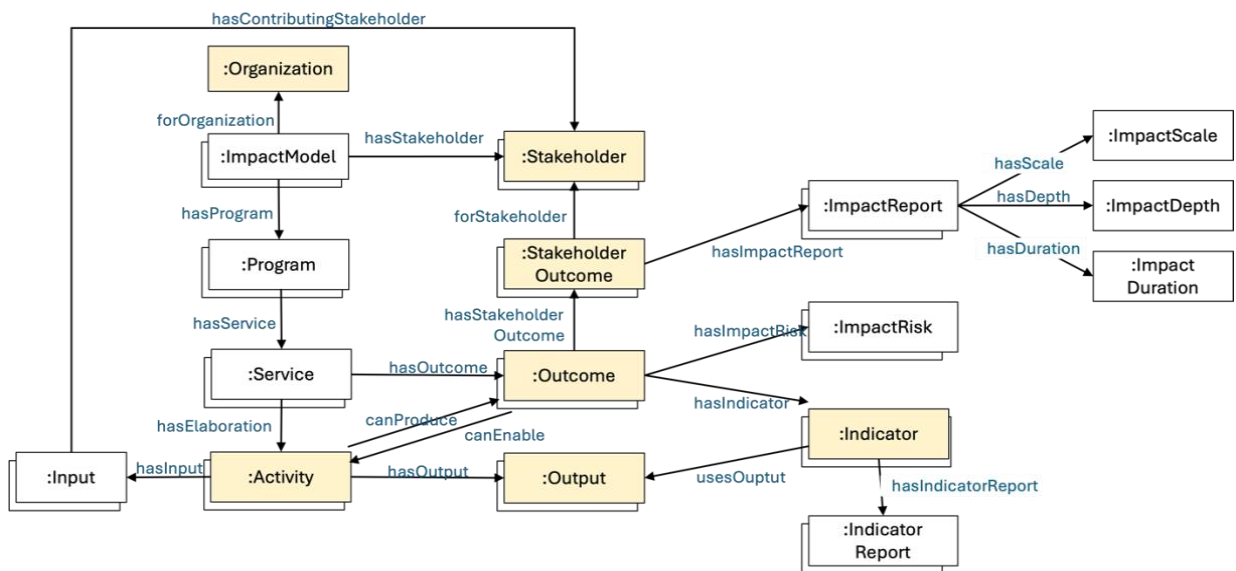

*Figure 11. Core classes of the Common Impact Data Standard used for modelling traceable impact.*

Figure 11 highlights the core classes needed to represent the traceable impact of activities in a food supply chain. These include:
- **Organization**: The organization who's impact is being traced through its activities.
- **Stakeholders**: Organization or Person type related to the outcome of an activity.
- **Characteristic**: A taxonomy of codes or codes that identify the stakeholder.
- **Activity**: An action that occurs in the domain and can be enabled by or causes some Outcome.
- **Output**: A quantitative summary of an activity.
- **Outcome**: What stakeholders experience as a result of activities.
- **Indicators**: A metric used to measure the outcome of activities, including a location, a time interval, a value, and a unit of measure.
- **Indicator Report**: Reports the value of an indicator for some time interval.
- **Impact Report:** Specifies the "How Much" dimension of the Impact Management Project. It reports on the scale, depth, and duration of the Outcome.
- **Impact Risk:** Assesses the likelihood that impact will be different than expected and that the difference will be material from the perspective of people or the planet who experience impact.

Note that CIDS incorporates the 5087-1 ontologies for activities, resources and time.

# Integrating Impact Modelling and Traceability

We align the 5087-1 Activity and Resource ontology with TOVE's traceability ontology and CIDS indicators to construct a model for tracing impact. While CIDS provides a comprehensive model for representing activities and impact, TRUs provide a model to connect activities and resources at time points and locations. Consider the scenario outlined in Figure 13 to Figure 21. We are interested in the claim that food delivered to consumers is produced locally, where the definition of locally produced goods is anything that was produced within 1000km from the place of use or consumption.

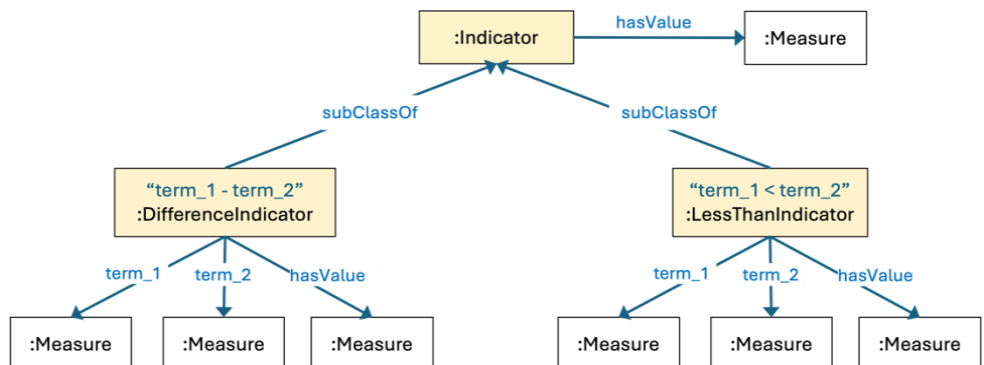

*Figure 12. :Indicator subclasses :DifferenceIndicator and :LessThanIndicators.*

To determine whether food was used or consumed within 1000km of production, we rely on two indicators to perform the calculation, as per Figure 12, where ":" denotes the type of the instance. Both classes are subclasses of the i72:*Indicator* class for which has the hasValue property sottres the indicator's value as an instance of a :*Measure* class. The :*DifferenceIndicator* takes two instances of the :*Measure* class as *term_1* and *term_2*, and calculates the difference between the two, resulting in a new instance a :Measure value. The :*LessThanIndicator* class also takes two instances of :Measure class as *term_1* and *term_2*, and stores "True" as the indicator value :Measure if term_1>term_2, and "False" otherwise.

In the following scenario, a meat packing plant in Guelph, ON produces 200lbs of beef, which is transported 664 km to a grocery store in Montreal, QC where it is purchased by consumers. Figure 13 illustrates an alignment of a CIDS :*LessThanIndicator* for "locally produced beef" and the :*ProduceState* :tru to measure the impact of the "Beef Production" activity.

The organization "Westcoast Meat" is an instance of class :*MeatPacker*, and the CIDS class :*Organization*. It performs the activity of producing beef ("Beef Production"), which is an instance of the 5087-1 :*Activity* class. At time $T_0$ the organization performs the "Beef Production" activity which produces 200 lbs of beef at their plant in "Guelph, ON." The activity has the effect of creating the :*ProduceState*, an instance of :tru, at the time point $T_0$, signifying the creation of 200 lbs of beef at location "Plant, Guelph, ON."

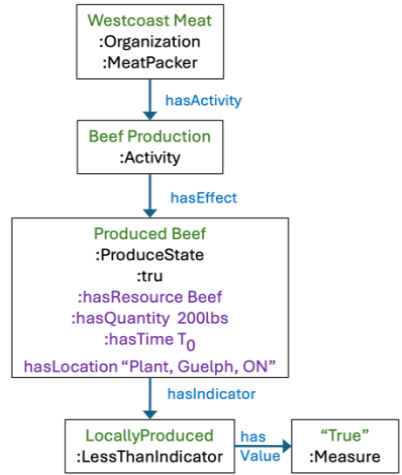

Figure 13. "ProduceState" TRU example: produce ground beef patties, with an indicator for "locally produced".

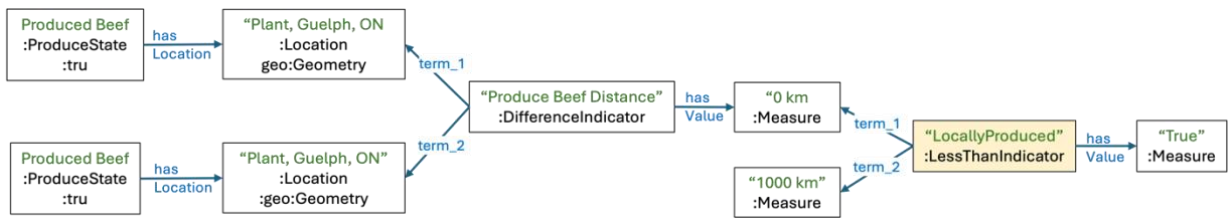

Figure 14. Calculation of the "LocallyProduced" indicator up to the "Produce Beef" :tru.

While trivially true, Figure 14, illustrates the calculation of the "LocallyProduced" Indicator for this :*tru*. The :*tru* has the location "Plant, Guelph, ON" as the location of production. First, the :*DifferenceIndicator* calculates the differences between two values. Since both are "Plant, Guelph, ON", the result is a :*Measure* instance with value "0 km". Next, the "LocallyProduced" indicator take two values, the previously calculated value of "0 km" and the threshold for which the indicator is true, namely "1000 km". For the "Produced Beef" :*tru* in Figure 14, the "LocallyProduced" indicator is true, since "0 km" < "1000 km".

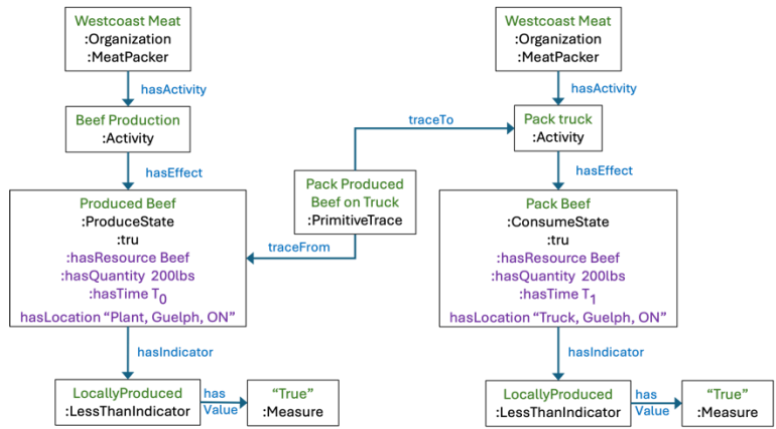

Figure 15. Tracing a TRU example: after the production of beef patties, they are packed onto a truck.

Figure 15 illustrates the next activity's :*tru*, namely the packing of produced beef onto a truck. The packing of the beef onto the truck, "Pack Beef", is an instance of the :tru class :*ConsumeState* since it consumes the existing resource, 200 lbs of beef, at time point $T_1$. The location has changed from "Plant, Guelph ON" to "Truck, Guelph, ON." This example illustrates the linking of two :tru's using the :*PrimitiveTrace* class, which links a :tru to an :*Activity*. In Figure 15, the "Pack Produced Beef on Truck" :*PrimitveTrace* is a trace from "Price Beef" :*tru* to the next activity, "Pack Truck" and its :*tru*, "Pack Beef".

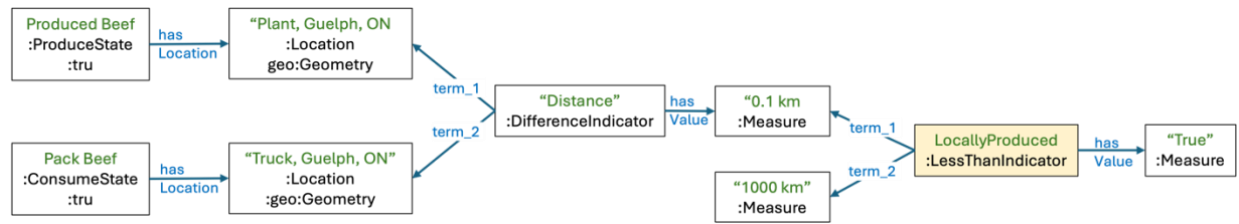

*Figure 16. Calculation of the "LocallyProduced" indicator up to the "Pack Beef" :tru.*

In Figure 16, the locations of "Produced Beef" :*tru* and "Pack Beef" :*tru* are used to calculated the difference indicator with a value of "0.1 km". Given that the beef was packed onto a truck 0.1 km away, it also satisfies the claim that the beef is locally produced, and the "LocallyProduced" indicator has a value of "True."

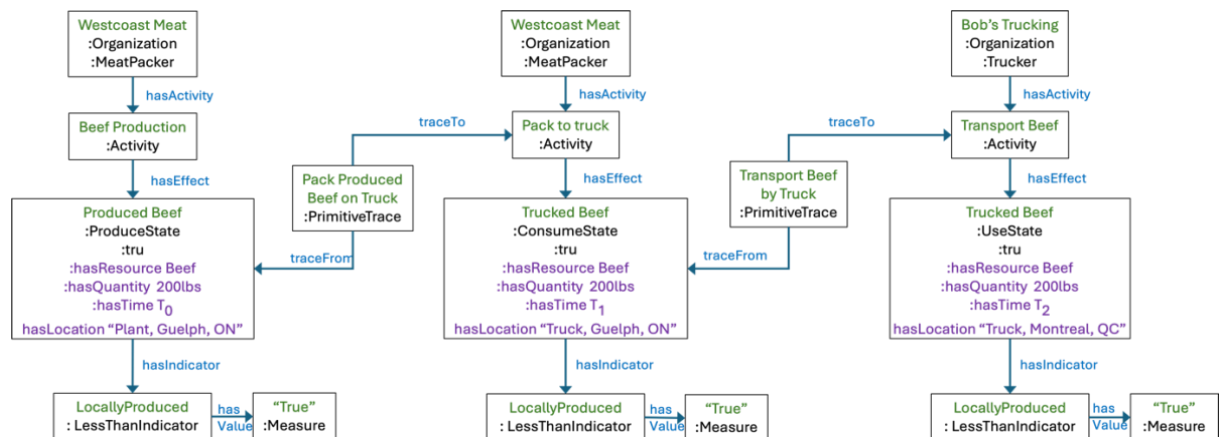

*Figure 17. Tracing a TRU example: after production of beef patties and packing them on a truck, they are transported by a trucking organization.*

Figure *17* illustrates the next activity's :*tru*, namely the transport of the produced beef. The organization "Bob's Trucking" is an instance of both the *:Trucker* class and *:Organization* class. It transports beef from "Truck, Guelph, Ontario" to its final destination, namely "Truck, Montreal, Quebec." The "Trucked Beef" :*UseState* :*tru* transition's the beef's state from "Produced Beef" to "Trucked Beef" since it uses the resource beef without consuming any of it, creating a new :*tru* at time-point $T_1$. Given that the transport's destination is within 1000 km, it also satisfies the claim that the beef is locally produced, and the "LocallyProduced" indicator has a value of "True."

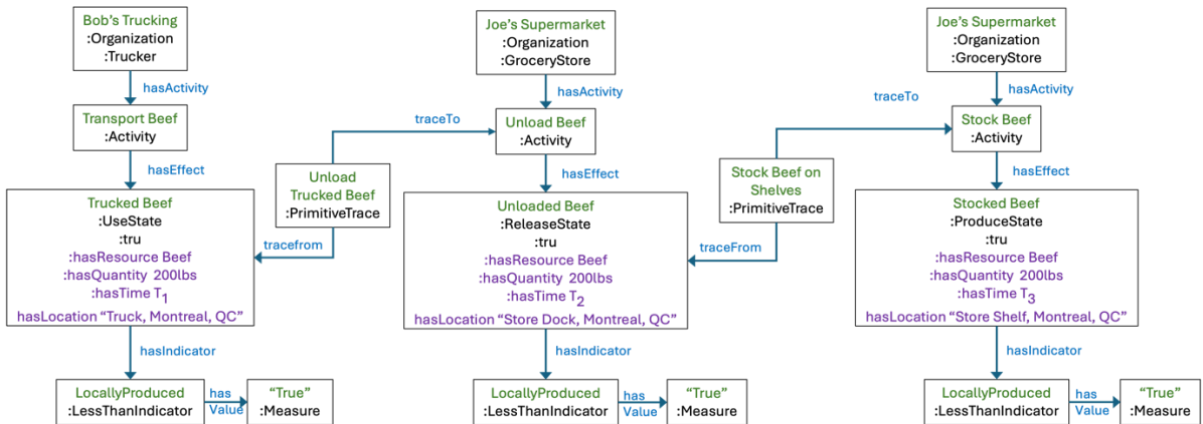
Figure 18. Tracing a TRU example: producing beef patties and transporting them to Montreal.

Figure 18 illustrates the next step in the supply chain, namely the delivery of the beef to the grocery store 644 km from the meat packer. At time point $T_2$, the grocery store organization "Joe's Supermarket" receives the beef and stocks it in their store. "Joe's Supermarket" is an instance of both the *:GroceyStore* class and *:Organization* class. It performs the activity of "Stock Beef", creating the *:ProduceState* *:tru* at $T_2$ with 200 lbs of beef. Again, since the location of the "Stocked Beef" *:tru* is within 1000 km of the meat packer, this beef is locally produced, and the "LocallyProduced" indicator's value is "True" at time-point $T_2$.

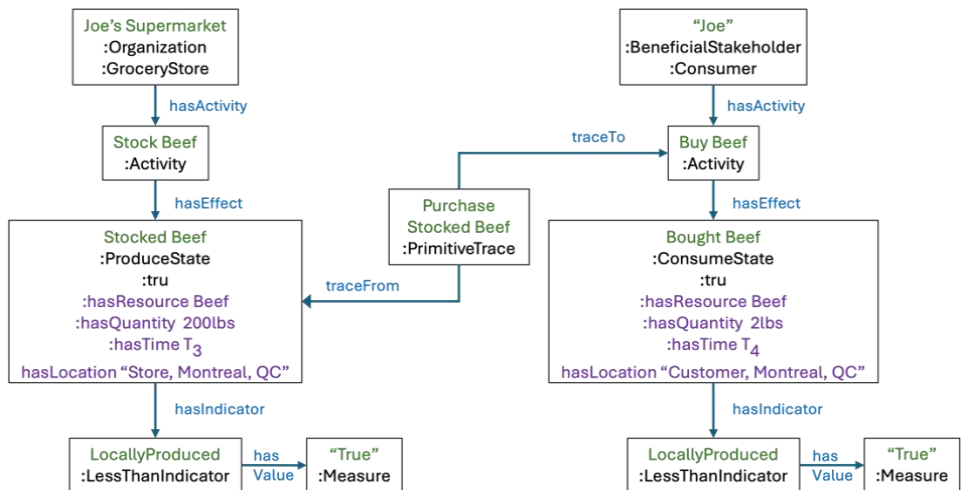
Figure 19. Example: Traceable Impact for assessing the impact lack of food has on the consumer.

The final step in the supply chain is the selling of beef to consumers. Figure *19* illustrates this by tracing the buying of 2 lbs of beef by a consumer at time point $T_4$. The consumer "Joe", is an instance of the *:Consumer* class as well as a member of the CIDS class *:BeneficialStakeholder* that identifies a group of stakeholders that benefit from resources in the supply chain. The consumer performs the activity of "Buy Beef" in the amount of 2 lbs at time-point $T_3$. This activity transitions a portion of the 200 lbs of beef, namely 2 lbs, from "Stocked Beef" to "Bought Beef," creating a *:ConsumeState* *:tru* at time point $T_4$ for 2lbs of the beef shipment.

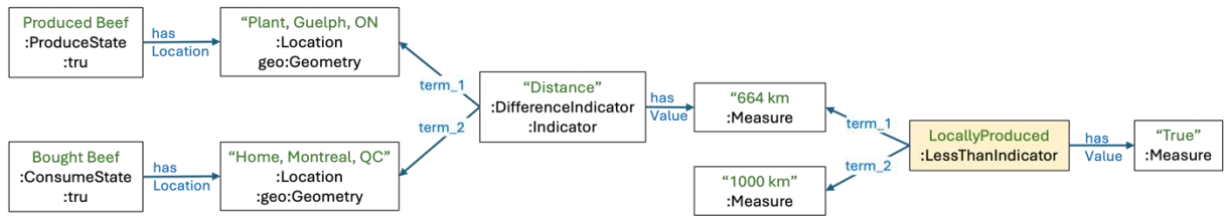

*Figure 20. Calculation of the "LocallyProduced" indicator for the packing of the "Bought Beef" :tru.*

In Figure 20, the "LocallyProduced" indicator is calculated from the point of production to point of consumption by the customer. The locations of "Produced Beef" :*tru* ("Plant, Guelph, ON") and "Bought Beef" :*tru* ("Home, Montreal, QC") are used to calculated the difference indicator with a value of "644 km". Given that the beef was bought 644 km away from being produced, it satisfies the claim that the beef is locally produced, and the "LocallyProduced" indicator has a value of "True" at $T_4$.

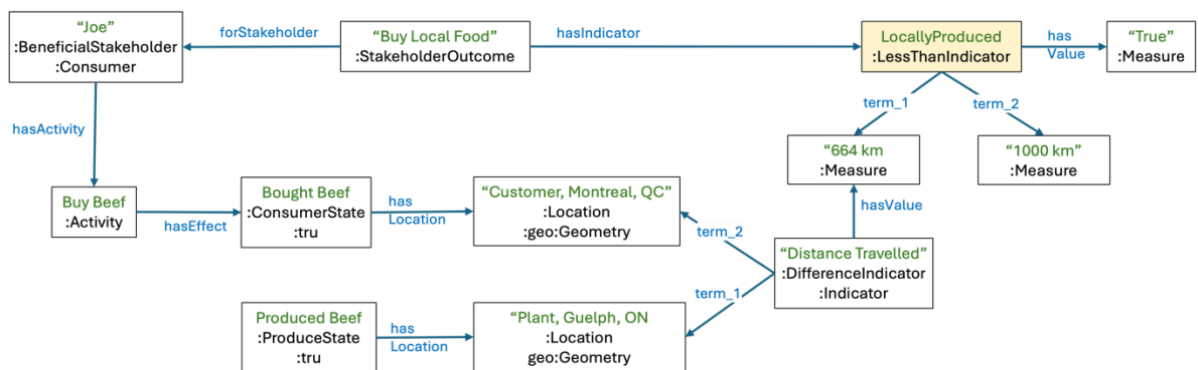

*Figure 21. Example: Traceable Impact for assessing the impact lack of food has on the consumer.*

Finally, in Figure 21, we align the newly calculated "LocallyProduced" indicator with the outcome consumer "Joe" is interested in. The consumer's :*Activity* instance "Buy Beef" has an effect, namely the :*ConsumeState* :*tru* of "Bought Beef." An instance of the :*StakeholderOutcome* class represents the outcome the consumer is interested in, namely "Buy Local Food." The indicator "LocallyProduced" is then linked to the stakeholder outcome through the *hasIndicator* property. First, we calculate the "Distance Travelled" Indicator between location of the two :tru's, giving "644 km". Next, "Locally Produced" calculation is based on the calculated distance "644 km" and threshold of "1000 km", giving "True" for the "LocallyProduced" indicator value.

## Discussion

The scenario in the previous section illustrates the model's ability to trace the claim of local production as it pertains to the production, transport, sale, and consumption of beef. The model could be extended beyond sales and identify other impacts on the consumer and beyond, such as the environment. For example, the model can trace how a consumer's nutrition levels are impacted. With the use of counterfactuals, we can identify the risk of, say, not delivering beef to grocery stores on the consumer's nutritional intake of protein and other nutrients. In our scenario, 200 lbs of beef would impact 100 consumers in the community. This shortfall must be reconciled by other means. Not doing so would pose a risk to the healthcare system by potentially causing malnutrition in parts of the population.

Equally, we can trace the environmental risk of food supply chain. For example, consider that the beef is not transported properly above, say 4.4°C, and must be thrown out. While the consumer is impacted by not consuming the beef, the environment is impacted by producing food waste, adding 60,000 kg of carbon dioxide and contributing to global warming. Meanwhile, the rotten beef and packs need to be dealt with separately due to different material properties. Rotten beef will be incinerated and continue to produce greenhouse gases, while the pack of plastic may go to a landfill, especially in some developing countries, thus creating further damage to the soil structure.

# Conclusion

A Traceable Impact Model requires understanding who the stakeholders are, their characteristics, the nature of their food insecurities, and their desired outcomes. With this understanding, we can work backwards through the food supply chain to identify where events and decisions adversely affect stakeholder outcomes. The result of such an investigation are recommendations aimed at re-engineering food supply chain processes. These recommendations are not merely corrective but are designed to address systemic shortcomings, thereby reinforcing accountability throughout the food supply chain.

An ontological impact model of the food supply chain serves as a comprehensive representational framework to address quality and other issues within the supply chain systematically. It establishes a logical foundation that identifies key concepts and their interrelations with quality assurance and control measures. This model conceptualizes traceability and the flow of data artifacts, allowing stakeholders to visualize and understand the movement and transformation of goods from production to consumption. This model's use of shared terminology and axioms is essential for ensuring coherence and consistency across various supply chain elements, facilitating clear communication and understanding among different parties. The axiomatization of the traceability model is forthcoming.

Furthermore, the ontological model forms the blueprints and foundation for developing traceable impact modelling software systems. Providing a structured and standardized schema for data enables the software to capture, process, and analyze information efficiently. This foundation is crucial for traceability systems to function effectively, as it underpins the software's ability to track the provenance, handling, transformation, and distribution of food products, thereby ensuring the integrity of the supply chain and facilitating responsiveness to quality issues, recalls, and regulatory compliance.